# Correlation between doping induced disorder and superconducting properties in carbohydrate doped $MgB_2$

J. H. Kim and S. X. Dou
*Institute for Superconducting and Electronic Materials, University of Wollongong, Northfields Avenue, Wollongong, NSW 2522, Australia*
Sangjun Oh[a)]
*Material Research Team, National Fusion Research Institute, 52 Eoeun-dong, Yuseong, Daejeon 305-333, Korea*
M. Jerčinović and E. Babić
*Department of Physics, Faculty of Science, University of Zagreb, HR-10000 Zagreb, Croatia*
T. Nakane, H. Kumakura
*Superconducting Materials Center, National Institute for Materials Science, 1-2-1 Sengen, Tsukuba, Ibaraki 305-0047, Japan*

A comprehensive study of the effects of carbohydrate doping on the superconductivity of $MgB_2$ has been conducted. In accordance with the dual reaction model, more carbon substitution is achieved at lower sintering temperature. As the sintering temperature is lowered, lattice disorder is increased. Disorder is an important factor determining the transition temperature for the samples studied in this work, as evidenced from the correlations among the lattice strain, the resistivity and the transition temperature. It is further shown that the increased critical current density in the high field region can be understood by a recently-proposed percolation model [Phys. Rev. Lett. 90 (2003) 247002]. For the critical current density analysis, the upper critical field is estimated from a correlation that has been reported in a recent review article [Supercond. Sci. Technol. 20 (2007) R47], where a sharp increase in the upper critical field by doping is mainly due to an increase in lattice disorder or impurity scattering. On the other hand, it is shown that the observed reduction in self-field critical current density is related to the reduction in the pinning force density by carbohydrate doping.

## I. INTRODUCTION

One of unique features observed in the superconductivity of $MgB_2$ is its two band nature: the almost isotropic $\pi$ bands originate from $p_z$ orbitals of the boron atoms and the highly anisotropic $\sigma$ bands from the in-plane $p_{xy}$ orbitals. The transition temperature can be affected not only by changes in the electronic density of states, or hardening or softening of phonons, but also by variations in the $\sigma$ gap anisotropy or interband scattering. It has been argued that high field behavior is dominated by the $\sigma$ bands [1], and as a result, the upper critical field of $MgB_2$ is reported to be anisotropic as well. Nevertheless, although there is a controversy on whether it is mainly due to interband or intraband scattering, the upper critical field can be enhanced significantly with increased impurity scattering, for example, by carbon doping [2]. It is now well established that carbon substitutes boron at the boron lattice sites and that the reduction in the transition temperature is rather small compared with the substantial increase in the upper critical field. It has been further claimed that the doping can enhance the pinning properties, for example, by SiC nanoparticle doping [3]. In this respect, the percolation model recently proposed by Eisterer and his co-workers [4] is of interest, because the critical current density is fitted using the intrinsic superconducting parameters, the upper critical field along the *ab*-plane, $B_{c2}^{ab}$, and the anisotropy parameter, $\gamma$. It was shown that the field and temperature dependence of the critical current for bulk and wires, irradiated or not, can be described by only four parameters including two extrinsic fitting parameters, the pinning force maximum, $F_m$, and the percolation threshold, $p_c$ [4]. We recently reported that even the strain dependence of the critical current can be understood by the percolation model as related to a variation in the anisotropy parameter [5].

A comprehensive study can be beneficial to elucidate the effects of doping on the superconductivity of $MgB_2$. SiC doping has been limited by local agglomerations of unreacted SiC or decomposed carbon and in this work, the effect of carbohydrate doping via the chemical solution method has been studied. Changes in the lattice parameters are discussed with a focus on lattice strain and its relation to the transition temperature. A correlation between the transition temperature, the upper critical field, and the critical current is further discussed.

## II. EXPERIMENTAL

We have previously reported on carbohydrate (malic acid, $C_4H_6O_5$) doping, which enables highly uniform mixing as compared with the solid state reaction method [6]. $MgB_2$ + 10wt% $C_4H_6O_5$ powder was prepared by using a chemical solution route: $C_4H_6O_5$ powder dissolved in toluene solvent form slurry with boron powder, the solvent is dried out, and the boron powder surface is encapsulated with carbon. Since the magnetically determined critical current is significantly lower

a)Electronic mail: wangpi@nfri.re.kr

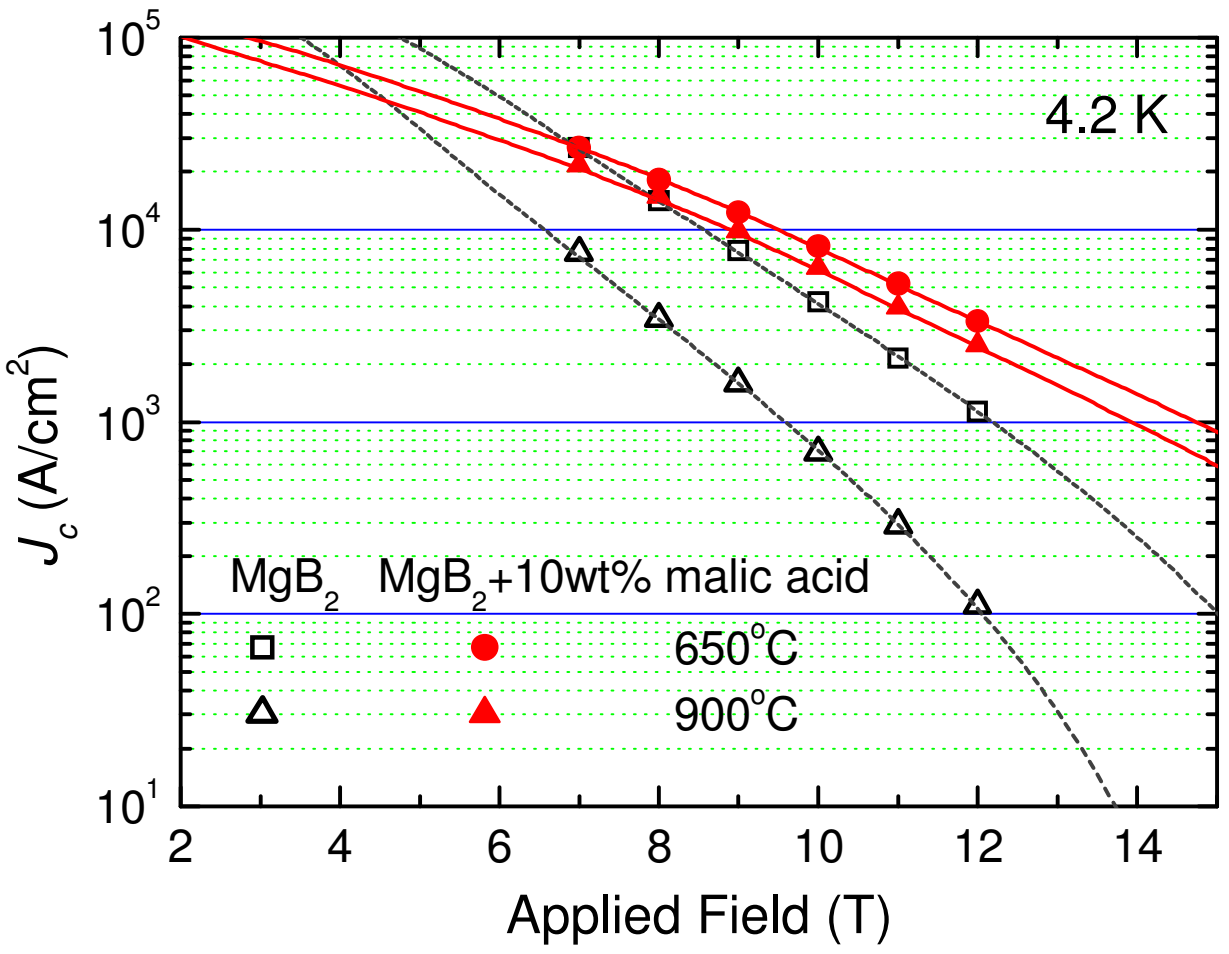

Fig. 1. The critical current density as a function of applied field for both un-doped and malic acid doped $MgB_2$ samples. Lines were calculated with the percolation model.

than the critical current measured by the transport method for $MgB_2$ [7], samples were fabricated in the form of wires with an iron sheath [8]. As a measure of impurity scattering, the resistivity was measured after removing the iron sheath and compared.

## III. RESULTS AND DISCUSSION

The critical current density measured at 4.2 K under various applied fields is shown in Figure 1. Consistent with previous reports, there is an obvious enhancement in the critical current density due to carbon doping at high fields, above 8 T. For both samples, doped and un-doped, the critical current density was higher at lower sintering temperature. This is opposite to what is usually observed for nano-carbon doped samples [3]. According to the dual reaction model, $MgB_2$ formation and carbon incorporation can occur simultaneously at temperatures as low as 600 °C if fresh reactive carbon is available [3]. The substitution of carbon into boron sites at low sintering temperature can be more clearly and directly seen from the variation in the lattice parameters. Interestingly enough, the *a*-axis lattice parameter for the carbohydrate doped sample was slightly increased as the sintering temperature was raised, as can be seen in Figure 2(a). The change in the lattice parameters can be converted into an actual amount of carbon substitution by comparing with the results on single crystals [9]. As shown in Figure 2(c), more carbon incorporation is achieved at a lower heat-treatment temperature for carbohydrate doping. Even for SiC doping, a slight increase in the carbon substitution level with increasing sintering temperature has been reported [3]. Low sintering temperature is usually regarded as being beneficial for flux pinning, due to the increase in the density of grain boundaries [10].

The $\sigma$ bands are known to be strongly coupled with the optical $E_{2g}$ phonon mode, the boron bond in-plane stretching mode [1], and changes in the lattice parameters or the unit cell volume can influence the transition temperature as shown in Figure 2(b). There can be stiffening in the $E_{2g}$ phonon mode or changes in the electronic density of states. However, it should be noted that the change in the lattice parameters is negligible for the $MgB_2$ sample without doping whereas there is a notable change in the transition temperature as a function of sintering temperature. For the SiC doping case, an opposite trend was observed: the transition temperature increased and the lattice parameter decreased with increasing sintering temperature [3].

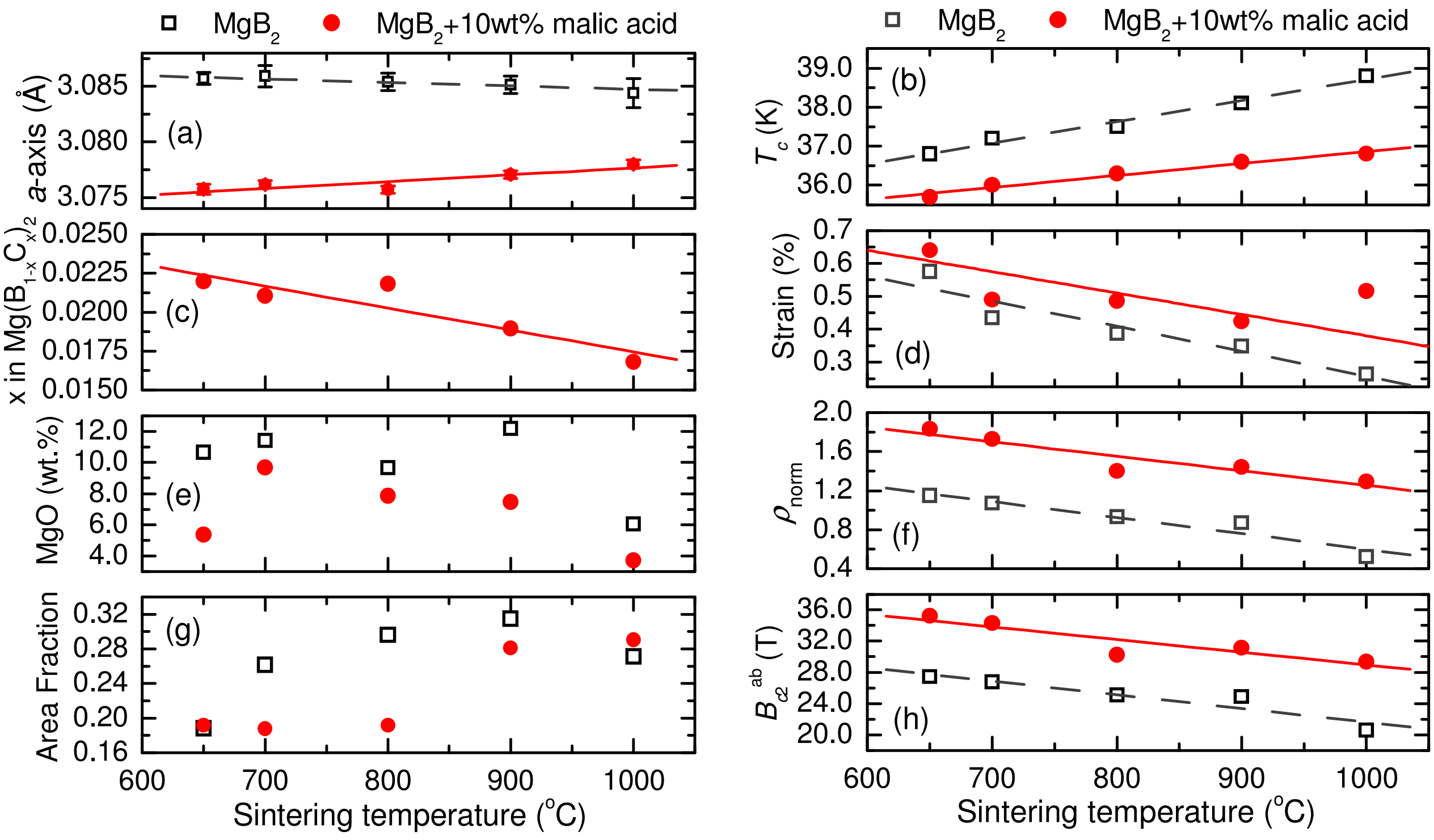

Fig. 2. (a) The *a*-axis lattice parameter, (b) the transition temperature, (c) the amount of carbon substitution, (d) the estimated lattice strain, (e) MgO weight percentage, (f) the normalized resistivity, (g) the area fraction, and (h) the estimated upper critical field at various sintering temperatures. All lines are only guides to the eyes.

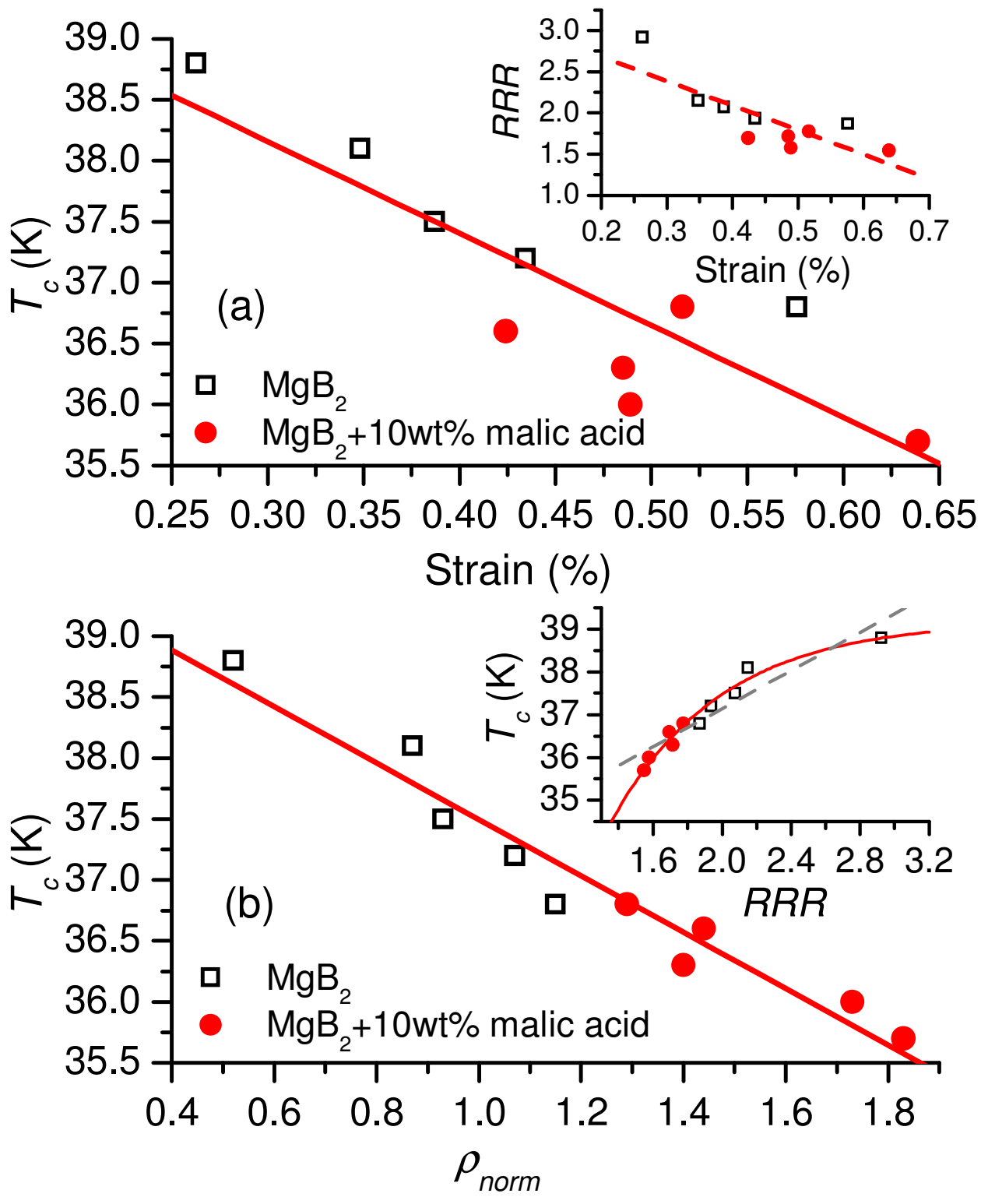

Fig. 3. (a) The transition temperature and the residual resistance ratio (inset) as a function of estimated lattice strain. Lines are only guides to the eyes. (b) The transition temperature as a function of the normalized resistivity and the residual resistance ratio (inset). Solid lines are calculated with $T_c = (39.81 - 2.314\ \rho_{norm})$ K and $T_c = 39.2\cdot(1 - \exp(-RRR/0.64))$ K, respectively.

On the other hand, a closer correlation can be found between the transition temperature and the lattice strain estimated from a Williamson-Hall plot using the isotropic model [11], as can be seen in Figure 2(d) or in Figure 3(a). At all sintering temperatures, the lattice strain of the carbon doped sample was higher than that of the pristine sample. Oxygen detached from the malic acid during the heat-treatment, which, in turn, can generate more MgO precipitates within the $MgB_2$ grains, can be a cause of the larger lattice strain observed for the malic acid doped samples. The MgO fraction obtained from Rietveld refinement as a function of sintering temperature is shown in Figure 2(e). Quite interestingly, the MgO fraction for the carbohydrate doped samples is lower than that of the pure samples at all temperatures. At least under our empirical conditions, it can be argued that MgO precipitates are not the major cause of the lattice strain. Other structural defects, such as vacancies, interstitials, substitutions or stacking faults [11], are other possible reasons for the observed lattice strain. It has been reported that there is a correlation between the resistivity and the lattice strain [12], as can be seen in the inset of Figure 3(a). Resistivity is affected by carrier scattering both at grain boundaries and within grains. Lattice strain can be a good measure for quantifying disorder within grains.

If disorder, which results in increased interband or intraband impurity scattering, is an important factor affecting the transition temperature of $MgB_2$ samples, there should be a relation between the resistivity and the transition temperature. In an early review, the Testardi correlation [13], an empirical relation between the residual resistivity ratio (*RRR*) and the transition temperature, was discussed [14]. As shown in the inset of Figure 3(b), we found that the transition temperature shows a dependence on the residual resistivity ratio (*RRR*), which can be by a simple exponential function: $T_c$ = 39.2·(1 - exp(-*RRR*/0.64)) K (solid line in the inset of Figure 3(b)). A more interesting correlation is suggested in a recent review of Eisterer [1] between the transition temperature and the normalized resistivity, the ratio of the residual resistivity to the difference in resistivity between 300 and 40 K, $\rho_{norm} = \rho_0/\Delta\rho$, where $\Delta\rho = \rho(300\text{ K}) - \rho(40\text{ K})$. Various reported data were compared, and it was argued that the transition temperature can be described by a simple linear relation, $T_c = (39.43 - 2.515\ \rho_{norm})$ K [1]. The sintering temperature dependence of the normalized resistivity is shown in Figure 2(f), and the correlation can be clearly seen in Figure 3(b). We found that our data are also in agreement with the above linear relation and can be better fitted with slightly different coefficients: $T_c = (39.81 - 2.314\ \rho_{norm})$ K. Our results support the idea that the transition temperature of $MgB_2$ is more closely related to the level of impurity scattering.

The above relation between the transition temperature and the normalized resistivity is of interest because it was further claimed that the zero temperature upper critical field along the *ab*-plane direction can be written as a function of $T_c$ and $\rho_{norm}$ [1]:

$$B_{c2}^{ab}(T=0) = \left(13.8\left(\frac{T_c}{T_{c,0}}\right)^2 + 14.7\frac{T_c}{T_{c,0}}\rho_{norm}\right)\text{ T}, \qquad (1)$$

where, $T_{c,0}$ is the maximum transition temperature from the above linear relation between $T_c$ and $\rho_{norm}$. The relationship was suggested by the well known relations between the upper critical field and the Ginzburg-Landau parameter $\kappa$, $B_{c2} = \sqrt{2}B_c \cdot \kappa$ and $\kappa = \kappa_{pure}\cdot(1+\xi_0/l)$ on the assumptions that the Bardeen, Cooper, and Schrieffer (BCS) coherence length $\xi_0$ is almost proportional to the transition temperature, $B_{c2}^{clean} = \sqrt{2}B_c \cdot \kappa_{pure} \propto \xi_0^{-2}$, and that the normalized resistivity is nearly inversely proportional to $l$, the electronic mean free path. Even though the effect of $\sigma$ bands carrier scattering on the upper critical field and that of $\pi$ bands scattering can be totally different [15], it was not considered in this work. Quite interestingly, it was shown that equation (1) above is in good agreement with empirical data reported in the literature, if the normalized resistivity is less than 2 [1]. For all the samples measured in this work, the normalized resistivity is less than 2, and the upper critical field along the *ab*-plane calculated from equation (1) is shown in Figure 2(h). $T_{c,0}$ of 39.81 K was used for the estimation. The enhancement of $B_{c2}^{ab}$ at low sintering temperature and with carbohydrate doping is mainly due to the increase in the normalized resistivity even though it causes a slight decrease in the transition temperature. If there is an enhancement in the upper critical field as shown in Figure 2(h) due to doping, obviously there can be an increase in the critical current density as well especially at high field as shown in Figure 1.

The critical current density shown in Figure 1 can be fitted reasonably well by a recent percolation model (lines in Figure 1) using the upper critical field shown in Figure 2 (h). The main idea of the percolation model is to calculate the overall critical current numerically based on a percolation theory [4]:

$$J_c = \int_0^\infty \left( \frac{p(J) - p_c}{1 - p_c} \right)^{1.79} dJ \qquad (2)$$

where $p(J)$ is the fraction of grains having their critical current density above $J$ and $p_c$ is the minimum fraction needed for a percolative superconducting current flow. The critical current of each grain is assumed to be described by the same pinning model, for example, by a grain boundary pinning model: $J_c = F_m \cdot (1 - B / B_{c2})^2 \big/ \sqrt{B_{c2} B}$ . It is assumed that the pinning force maximum, $F_m$, of each grain is the same. The difference in the critical current density for each grain is due to the upper critical field anisotropy, which can be described by: $B_{c2}(\theta) = B_{c2}^{ab} \big/ \sqrt{\gamma^2 \cos^2(\theta) + \sin^2(\theta)}$ , following the anisotropic Ginzburg-Landau theory. The best fit (lines in Figure 1) is obtained with the anisotropy parameter $\gamma$ and the pinning force maximum $F_m$ values that are listed in Table I. As has been mentioned, the upper critical field shown in Figure 2(h) was used for the fitting. The percolation threshold $p_c$ of ~0.2 for densely packed $MgB_2$ bulk samples and around 0.26 for wire samples has been reported [4], and $p_c$ of 0.26 was used in this work.

One of the interesting features observed in Figure 1 is that the decrease in the critical current density with higher sintering temperature is significantly lower for the carbohydrate doped samples. From the obtained parameters, it can be argued that this is due to the decrease in the anisotropy parameter with increased sintering temperature for the malic acid doped sample. However, it should be noted that the critical current density in the high field region can be affected either by the change in the anisotropy parameter or by the variation of the upper critical field. The parameters used are comparable to previous reports, and the self-consistency was checked by comparing the irreversibility field, $B_{irr}$, obtained from the above critical current fitting results and the estimation from the resistive broadening: $B_{irr} = B_{c2}^{ab} \big/ \sqrt{(\gamma^2 - 1) p_c^2 + 1}$ [16], but even so, there is some uncertainty in the determination of the parameters.

Table I
The fitting parameters used for the calculation from the percolation model, equation (2), of the critical current density shown in Figure 1.

| | Without doping | | With carbohydrate doping | |
|---|---|---|---|---|
| | 650 °C | 900 °C | 650 °C | 900 °C |
| $\gamma$ | 3.92 | 4.6 | 3.35 | 2.92 |
| $F_m$ (N/m³) | $2.85 \cdot 10^{10}$ | $1.95 \cdot 10^{10}$ | $1.1 \cdot 10^{10}$ | $0.85 \cdot 10^{10}$ |

The low field region is somewhat different. The critical current density at low field is mostly determined by the pinning force maximum and the obtained $F_m$ listed in Table I is quite reliable. Consistent with the conventional argument concerning the size dependence of the grain boundary pinning, there is a clear enhancement in the pinning at low sintering temperature for both doped and un-doped samples. On the other hand, contrary to the usual discussion on carbon doping in $MgB_2$, the pinning force did not increase, but decreased. As a simple measure of the effective cross-section for current transport, the area fraction ($AF$) has been proposed, where $AF = \Delta\rho_c / \Delta\rho$ [17]. $\Delta\rho_c$ is the reference phonon contribution to the resistivity for a clean sample and is about 9 μΩ·cm [1]. It has been recently suggested that the pinning force maximum should be modified as well, to include the effect of the reduced area: $F_m^* = F_m / AF$ [1]. The area fraction at various sintering temperatures is shown in Figure 2(g). Even if we include the effect of the area fraction, the above conclusion does not change: the pinning force decreases with increasing sintering temperature and by malic acid doping. We, therefore, conclude that the observed low self-field critical current density for the malic acid doped sample is due to the reduced pinning force and suggest that further improvement in the pinning force is still crucial for the development of $MgB_2$ wire for engineering applications.

## IV. CONCLUSIONS

In summary, it was argued that disorder is an important factor affecting the transition temperature of $MgB_2$. There were close correlations among the lattice strain, the resistivity and the transition temperature for both un-doped and malic acid doped $MgB_2$ samples. Specifically, we found a linear relation between the normalized resistivity and the transition temperature: $T_c = (39.81 - 2.314\, \rho_{norm})$ K, as has been recently suggested. The upper critical field was estimated further from the relation reported: $B_{c2}^{ab} = 13.8\, (T_c/T_{c,0})^2 + 14.7\, (T_c/T_{c,0})\, \rho_{norm}$, which is in accordance with the empirical results that the reduction in $T_c$ due to carbon doping is small compared with the substantial increase in $B_{c2}^{ab}$ . The critical current density can be well described by the percolation model using the estimated upper critical field. Consistent with the dual reaction model, more carbon was substituted at lower sintering temperatures, which were as low as 650 °C for malic acid doping. Low sintering temperature results in increased disorder, but the transition temperature was decreased only down to 35.7 K, whereas the upper critical field is expected to be raised as high as ~35 T. The increased upper critical field and the reduced anisotropy parameter are likely to be reasons for the observed critical current density increment at high field for the doped sample. On the other hand, the low self-field critical current density for the malic acid doped samples compared with the un-doped samples is due to the reduction in the pinning force density, so we conclude that further improvement in the pinning force is necessary for the development of $MgB_2$ wire for engineering applications.

## ACKNOWLEDGEMENTS

This work was supported by the Australian Research Council, Hyper Tech Research Inc., OH, USA, and Alphatech International Ltd, NZ. The work done at the National Fusion Research Institute was supported by a Korea Science and Engineering Foundation (KOSEF) grant funded by the Korean Government (MOST) (No. R01-2007-000-20462-0).